%% file: main.tex
\documentclass[sigconf,nonacm]{acmart}
\usepackage{diagbox}
\usepackage{graphicx}
\usepackage{caption}

\usepackage{multirow}
\AtBeginDocument{%
  }

\input{0-ACSAC-Minor-Revision/commands}

\begin{document}

%\title{Can Large Language Models Provide S\&P Advice?  Measuring LLMs' Ability to Refute S\&P Misconceptions} 
%\title{Can Large Language Models Provide S\&P Advice?  Measuring LLMs' Ability to Refute S\&P Misconceptions} 
\title{{\fontsize{11}{12}\selectfont \textnormal{Accepted to the Annual Computer Security Applications Conference (ACSAC), 2023}} \\[1ex] Can Large Language Models Provide Security \& Privacy Advice? Measuring the Ability of LLMs to Refute Misconceptions} 

\author{Yufan Chen}
\authornote{The authors Chen and Arunasalam contributed equally to this research.}
\email{chen4076@purdue.edu}
\affiliation{%
  \institution{Purdue University}
  \city{West Lafayette}
  \state{IN}
  \country{USA}
}

\author{Arjun Arunasalam}
\authornotemark[1]
\email{aarunasa@purdue.edu}
\affiliation{%
  \institution{Purdue University}
  %\streetaddress{P.O. Box 1212}
  \city{West Lafayette}
  \state{IN}
  \country{USA}
}

\author{Z. Berkay Celik}
\email{zcelik@purdue.edu}
\affiliation{%
  \institution{Purdue University}
  \city{West Lafayette}
  \state{IN}
  \country{USA}
}

\input{0-ACSAC-Minor-Revision/Content/abstract}

\maketitle
\pagestyle{plain}
\input{0-ACSAC-Minor-Revision/Content/intro}

\input{0-ACSAC-Minor-Revision/Content/background}

\input{0-ACSAC-Minor-Revision/Content/approach}

\input{0-ACSAC-Minor-Revision/Content/evaluation}
\input{0-ACSAC-Minor-Revision/Content/discussion}
\input{0-ACSAC-Minor-Revision/Content/related_work}

\input{0-ACSAC-Minor-Revision/Content/conclusion}
\section*{Acknowledgements}
We thank our anonymous reviewers and shepherd for
providing us with valuable feedback that helped improve our paper. 
This work is supported by startup funding from Purdue University.

\bibliographystyle{ACM-Reference-Format}
\bibliography{reference}

.
\appendix

\end{document}

%% file: 0-ACSAC-Minor-Revision/commands.tex
\newboolean{commentsOn}
\setboolean{commentsOn}{true}
\usepackage{graphicx}
\usepackage{subfigure}

\def\eg{{e.g.},~}
\newcommand{\shortsectionBf}[1]{\vspace{3pt}\noindent {\bf #1}}
\newcommand{\shortsectionBfn}[1]{\vspace{3pt} {\bf #1}}

\usepackage[scaled=.8]{beramono}

\usepackage[framemethod=TikZ]{mdframed}   
\usepackage{xcolor}
\usepackage{multirow}
\usepackage{soul}
\usepackage{threeparttable}

\usepackage{array}
\newcolumntype{P}[1]{>{\centering\arraybackslash}p{#1}}

\usepackage{microtype}
\newmdenv[backgroundcolor=gray!5,
    linecolor=gray,
    linewidth=1.5pt,
    roundcorner=5pt,
    skipabove=7pt,
    skipbelow=0pt,
    font=\normalsize
]{shadedboxed}

\renewenvironment{quote}
  {\small\list{}{\rightmargin=0.5cm \leftmargin=0.5cm}%
   \item\relax}
  {\endlist}

\DeclareRobustCommand*\circled[1]{\tikz[baseline=(char.base)]{ \node[shape=circle,draw,color=white,fill=black,inner sep=0.5pt] (char){#1};}}

%% file: 0-ACSAC-Minor-Revision/Content/abstract.tex
\begin{abstract}
Users seek security \& privacy (S\&P) advice from online resources, including trusted websites and content-sharing platforms. 
These resources help users understand S\&P technologies and tools and suggest actionable strategies.
Large Language Models (LLMs) have recently emerged as trusted information sources. 
%Problem and Gap
However, their accuracy and correctness have been called into question. 
Prior research has outlined the shortcomings of LLMs in answering multiple-choice questions and user ability to inadvertently circumvent model restrictions (\eg to produce toxic content). 
Yet, the ability of LLMs to provide reliable S\&P advice is not well-explored.

In this paper, we measure their ability to refute popular S\&P misconceptions that the general public holds.
We first study recent academic literature to curate a dataset of over a hundred S\&P-related misconceptions across six different topics. 
We then query two popular LLMs (Bard and ChatGPT) and develop a labeling guide to evaluate their responses to these misconceptions.
To comprehensively evaluate their responses, we further apply three strategies: query each misconception multiple times, generate and query their paraphrases, and solicit source URLs of the responses.

Both models demonstrate, on average, a $21.3$\% non-negligible error rate, incorrectly supporting popular S\&P misconceptions. 
The error rate increases to $32.6$\% when we repeatedly query LLMs with the same or paraphrased misconceptions. 
We also expose that models may partially support a misconception or remain noncommittal, refusing a firm stance on misconceptions.
Our exploration of information sources for responses revealed that LLMs are susceptible to providing invalid URLs ($21.2\%$ for Bard and $67.7\%$ for ChatGPT) or point to unrelated sources  ($44.2\%$  returned by Bard and $18.3\%$ by ChatGPT). 
Our findings highlight that existing LLMs are not completely reliable for S\&P advice and motivate future work in understanding how users can better interact with this technology.
\end{abstract}

%% file: 0-ACSAC-Minor-Revision/Content/intro.tex
\section{Introduction}
% Area
In recent years, large language models (LLMs) have emerged as the most prominent technology in natural language processing (NLP). 
These models, trained on vast amounts of data, possess rich embedded knowledge, allowing them to be easily applied to various downstream NLP tasks such as sentiment analysis, code generation, and question answering with minimal fine-tuning~\cite{Brown2020Learner, bang2023multitask, wei2021finetuned}. 

The advent of LLM agents such as ChatGPT~\cite{chatgpt} and Bard~\cite{bard} has resulted in widespread public interaction with LLMs, with end users leveraging web interfaces to engage with these powerful AI tools.
These chatbots are trained to interact with users in a conversational style and generate answers to users' questions by retrieving information from the models themselves.

The usability of these web interfaces has caused LLMs to gain great popularity and resulted in widespread use among the general public in a short timeframe. 
These LLMs provide a single interface for users to interact with, contrasting conventional search engines which require users to traverse various web pages. 
Thus, LLMs have emerged as a new trusted source of information.
Today, people interact with the LLMs to obtain information regarding health~\cite{chatgpt-medical-advice}, stock market advice~\cite{chatgpt-stock-advice}, and even job interview guidance~\cite{chatgpt-job-interview}. 
As LLMs gain further footing in users' everyday lives, they become prominent information sources in popular domains, including user security and privacy~(S\&P) advice.
Traditionally, people receive S\&P advice from friends, family, and online sources (\eg forums, social media, IT websites)~\cite{nicholson2019if, redmiles2016think}. 
However, recent LLM popularity as a trusted resource, in lieu of search engines, provides prime conditions for its adoption as an S\&P advice tool.

%% Problem
Concerningly, so far, their ability to provide quality expert S\&P advice has not been well-evaluated. 
Previous research has recently revealed that LLMs such as ChatGPT can generate untruthful information, hallucination (\eg provoking fake news articles or academic papers), and even toxic content~\cite{bang2023multitask, zuccon2023dr, shen2023chatgpt, si2022so, gehman2020realtoxicityprompts}. 
This raises concerns, as lay users lacking expertise may easily trust and be misled by the generated falsehoods as the information LLMs provide is often written proficiently and convincingly. 
%in their response 
Extending to this, improper S\&P advice can have severe consequences. 
For instance, users may be misled into believing specific strategies (\eg reusing strong passwords) are secure and privacy-preserving when they are not (and vice versa).
Similarly, inaccurate information on S\&P tools/technologies may lead to erroneous decisions (\eg exclusion due to lack of perceived S\&P  or adoption due to being overpromised of their S\&P capabilities).
Therefore, it is crucial to thoroughly evaluate LLMs' performance in providing reliable S\&P advice.

In this paper, to address this, we empirically aim to answer the following research question, 
\vspace{-1pt}
\begin{quote}
 \textbf{Are LLMs reliable in providing S\&P advice by correctly refuting user-held S\&P-related misconceptions?}
\end{quote}
\vspace{-1pt}
Our approach to answering this research question begins with an extensive literature survey on user-held S\&P misconceptions.
We query Google Scholar~\cite{googlescholar} with a list of S\&P topics and user-related keywords (\eg ``folk models'', ``user perception'') to retrieve academic studies that address people’s misconceptions about security and privacy technologies and strategies. 
We collect a total of \textasciitilde$400$ academic manuscripts through this method.
After removing irrelevant manuscripts, we extract over $500$ S\&P misconceptions, remove redundancies, and produce a corpus of $122$ S\&P publicly held misconceptions. 
These misconceptions are grouped into six different S\&P topics, \eg Web Security and Privacy, IoT/CPS.

Using our dataset, we evaluate the ability of LLMs to refute these misconceptions. 
We choose ChatGPT and Bard as they are among the most well-known and influential LLMs available to the public. 
We conduct four experiments designed to understand LLM's overall correctness, consistency, susceptibility to paraphrasing, and reliability in providing sources to refute misconceptions. 
%.  
%
First, we query each misconception once to evaluate the general effectiveness of models in refuting misconceptions. 
Second, we perform repeated queries (four times) for each misconception to assess LLMs' consistency in maintaining their stance towards a misconception.
Third, we leverage paraphrasing tools to generate four paraphrases for each misconception and ask the models once for each paraphrase, aiming to simulate the real-world scenario in which people may query the same misconception differently. 
Lastly, we query for the URL sources that influence the models' responses and evaluate their validity and reliability (whether the URLs exist and what information the websites they direct to provide).
After collecting all the responses, we develop a labeling guide to categorize them according to their stances. 
The categories include \textit{support}, \textit{negate}, \textit{partially support}, \textit{noncommittal} and \textit{unrelated}. 
Since all entries in our dataset are misconceptions, we consider \textit{negate} as the only correct answer and \textit{support} as the incorrect answer.
We analyze the source URLs using the Python HTTP library \texttt{requests}~\cite{requests} and Wayback Machine API~\cite{wayback_machine} to determine URL validity.
We further analyze domain relevance for each URL to verify if the website it points to is relevant to the misconception.

% Evaluation
%results
Our experiments reveal that, 
$(1)$~ChatGPT and Bard both exhibit non-negligible error rates that averaged to be $21.3\%$.
$(2)$~When repeatedly queried with the same question, Bard performs better than ChatGPT in keeping consistent in its answers. 
However, ChatGPT and Bard both show an increase in error rate (by $10.6\%$ and $4.1\%$) and a decrease in correctness (by $12.3\%$ and $8.2\%$). 
$(3)$~When presented with paraphrased questions, both models experience an additional increase in error rate (by $6.6\%$ and $9\%$ for ChatGPT and Bard) compared to those of repeated queries.
Additionally, Bard's advantage in maintaining consistency diminishes, with both models providing at least two different stances for an average of $44.7$\% of the misconceptions. 
When analyzing models' ability in refuting misconceptions across different categories, it is observed that $(4)$~ChatGPT and Bard perform poorly in S\&P law and regulation category, with error rates exceeding $40$\% in almost all experiments. 

In our source analysis, it is found that $(5)$~Bard is more likely to provide valid URLs ($78.8\%$ validity) compared to ChatGPT, which generates mostly forged URLs ($32.3\%$ validity).
$44.2\%$ of Bard and $18.3\%$ of ChatGPT's valid URLs are unrelated to the misconception's domain, \eg a response to a VPN misconception provides a URL to a website that does not mention VPNs.
We further analyze a sample of sources provided when the LLMs erroneously support misconceptions.
We found that an average of $10$\% of the valid URLs attributed to incorrect answers contain false information, while an average of $25$\% contains correct information that debunks the misconception, yet ChatGPT and Bard fail to recognize it.
Our empirical findings highlight how existing state-of-the-art LLMs are not completely reliable in providing accurate S\&P advice or sound sources.
Our study highlights the necessity for future work on how users interact with LLMs as a trusted information source and the need to further examine LLM's ability to provide expert advice in different domains.

In this work, we make the following contributions:
\begin{itemize}
\setlength\itemsep{0.5em}
    \item We curated a dataset of over a hundred  S\&P misconceptions held by users through a comprehensive literature survey.
    \item We extensively evaluated two popular LLMs' ability to provide S\&P advice by measuring their correctness, consistency, and susceptibility to paraphrasing.
    \item We analyzed URLs that LLMs provide when justifying their stance towards misconceptions, demonstrating their inability to provide reliable sources for S\&P information.
\end{itemize}

 \noindent Our artifacts, which include our misconception dataset, labeling guide, and results,  are made available~\cite {replication}.

%% file: 0-ACSAC-Minor-Revision/Content/background.tex
\section{Background and Motivation}

\shortsectionBf{Large Language Models.} 
LLMs are neural network models with billions (or more) parameters. 
These models are trained on a large corpus of internet-sourced data, including textual information and conversation data. 
Such training enables LLMs to generate human-like language and exhibit capabilities in zero-shot and few-shot learning. 
With minimal fine-tuning, LLMs can be easily applied to various NLP tasks such as sentiment analysis, code generation, question answering, and, more recently, engaging in coherent conversations with users~\cite{Brown2020Learner, vaithilingam2022expectation, chatgpt-blog}.

ChatGPT and Bard have emerged as two of the most well-known and widely-used LLMs. 
ChatGPT is a conversational variant of InstructGPT~\cite{ouyang2022training} and has been fine-tuned using Reinforcement Learning with Human Feedback (RLHF)~\cite{chatgpt-blog}. 
Bard is a conversational AI service based on Google's Language Model for Dialogue Applications (LaMDA)~\cite{thoppilan2022lamda, Bard-blog}. 
Both models have garnered significant attention and are recognized for their effectiveness.

Due to their extensive training, LLMs possess remarkable capabilities in answering a wide array of questions. 
Consequently, users increasingly rely on LLMs as sources of information. 
LLMs are leveraged to assist users with daily activities such as writing code~\cite{chatgpt-coding} and answering homework questions~\cite{homework-gpt}, to critical events/tasks such as job interviews~\cite{job-interview} and purchasing stocks~\cite{chatgpt-stock-advice}.

Users who previously turned to online avenues such as conventional search engines (\eg Google and DuckDuckGo) as advice sources have recently pivoted to LLMs~\cite{replace_search}.
LLMs provide users with an interface where they can clarify doubts on topics ranging from health~\cite{health-gpt} to relationship advice~\cite{dating-gpt} without visiting multiple websites.
These topics will likely grow, encompassing niche areas where users previously sought information/advice online.

\shortsectionBf{Security and Privacy Advice.}
In an effort to protect themselves from various S\&P threats, people often seek S\&P advice from offline interactions with friends and family and online spaces such as forums, social media, and search engines.
Here, users learn about password security and how to protect themselves from malware and may even educate themselves on S\&P tools/technologies such as malware, privacy, blockchain, and VPNs~\cite{redmiles2020comprehensive}.
The quality of S\&P advice, however, can be concerning, mainly when sourced from online resources that are not vetted for accuracy. 
Advice pointers may be unactionable or lack clear priorities.
This may make it difficult for individuals to determine which advice to follow and thus prevent users from effectively implementing security and privacy-enhancing strategies~\cite{redmiles2020comprehensive}.

S\&P advice may additionally be inaccurate and categorically false, and it may lead to implementing insecure and privacy compromising suggestions.
As LLMs emerge as new trusted sources of information, people may increasingly rely on them for assistance with their S\&P concerns or understanding of  S\&P concepts/tools. 
Therefore, it becomes crucial to evaluate the effectiveness of existing LLMs in providing advice in these domains to ascertain their reliability as sources of information.

%% file: 0-ACSAC-Minor-Revision/Content/approach.tex
\begin{figure}
    \centering
\includegraphics[width=\columnwidth]{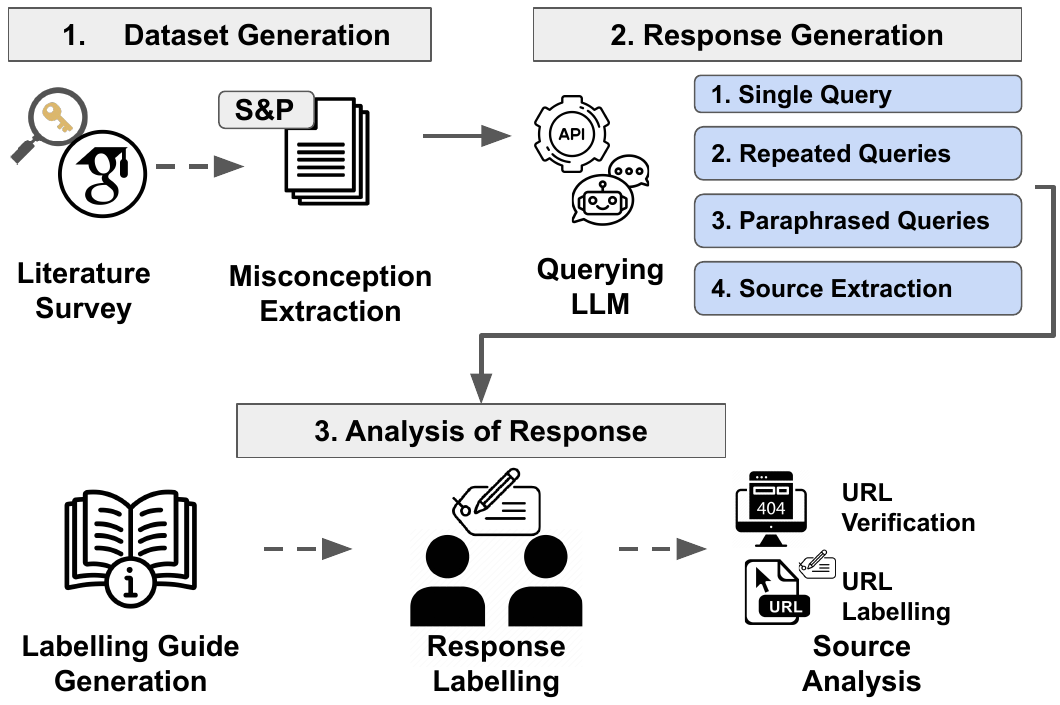}
    \caption{Overview of our methodology in understanding the ability of LLMs to refute S\&P misconceptions.}
    \label{fig:approach}
\end{figure}

\input{0-ACSAC-Minor-Revision/Content/tables/dataset-table}

\section{Methodology}
\label{sec:methodology}
We focus on empirically answering the following research question: \textbf{Are LLMs reliable in providing S\&P advice by correctly refuting S\&P-related misconceptions?}
Figure~\ref{fig:approach} presents an overview of how we address this research question. 
First, we curate a dataset of S\&P misconceptions the general public holds. 
To do so, we extensively study the literature on user misconceptions about S\&P topics and curate a dataset that contains diverse misconceptions covering a variety of topics, such as blockchain and malware.

Following a prompt template, we query two popular LLMs (ChatGPT and Bard) to collect data on how these LLMs respond to S\&P misconceptions. 
We then design a four-experiment approach to evaluate their ability to refute common S\&P  misconceptions. 
In an initial experiment, we query each LLM with all misconceptions in our dataset. 
Then for each misconception, we repeat our queries to obtain a total of five responses each.  
Thereafter, we leverage paraphrasing tools to generate semantically similar sentences for each misconception and repeat our queries. 
Lastly, we prompt LLMs for additional information, specifically asking them for the sources that informed their responses.

After responses are collected, two authors generate a labeling guide to label LLMs' responses independently and categorize them into five distinct groups (\textit{support}, \textit{negate}, \textit{partially support}, \textit{noncommittal}, \textit{unrelated}), and reconcile differences. 
We also analyze source URLs returned by the LLMs, leveraging both the Python HTTP library \texttt{requests}~\cite{requests}
and Wayback Machine API~\cite{wayback_machine} to verify URL's validity before further grouping URLs into one of three groups (\textit{relevant}, \textit{marginally relevant}, \textit{irrelevant}).
In the following, we detail each stage and present our findings in Section~\ref{sec:eval}.
We provide our replication package, which includes our dataset and all supplementary material necessary for reproducing our results~\cite{replication}.

\subsection{S\&P Misconception Dataset Generation}
We conducted an extensive literature survey on existing academic studies that detail general users' misconceptions about S\&P topics.
To do so, we first generated a comprehensive query list to query Google Scholar~\cite{googlescholar} using the template ``\{domain\} \{user-related keywords\}''.
For the ``\{domain\}'' portion of our query, we identified a comprehensive list of $14$  S\&P topics, \eg malware, web privacy, and cyber-physical systems (such as autonomous driving). 
For the ``\{user-related keywords\}'' portion, we used keywords that would encourage retrieval of studies focused on user perception, including keywords ``user study'', ``mental model'', ``folk model'', and ``user perception''. 
For instance, one query that we used in our retrieval was ``web privacy user perception.''

Overall, we generated $56$ queries with various combinations of ``\{domain\}''  and ``\{user-related keywords\}''.
Leveraging our query list, we gathered approximately \textasciitilde$400$ academic manuscripts consisting of diverse sources such as conference papers, journals, book chapters, and articles.

\shortsectionBf{Misconceptions Extraction and Filtering.} 
Two authors use a labeling guide to independently label manuscripts as relevant or irrelevant to S\&P misconceptions. 
First, authors read manuscript titles and remove unrelated papers.
They then read the abstract and introduction for further filtration. 
The authors marked a manuscript relevant if $(1)$~it discusses S\&P related misconceptions and $(2)$~notes that the general public holds these misconceptions.
Authors met to reconcile differences and achieved high agreement, Cohen's Kappa~\cite{fleiss1981measurement}, $\kappa > 0.8$, before reconciling.

After reconciling, we extracted \textasciitilde 500 misconceptions. 
However, after our initial analysis, we found that some misconceptions we extracted overlap. 
To address this, we manually filtered out duplicates to eliminate redundancy, resulting in a final dataset comprising $122$ distinct misconceptions.  
We note that our misconception dataset focused on recent publications; $121$ misconceptions were from publications between $2014-2023$, $1$ from $2006$.
Our full list of keywords, labeling guides, and misconception dataset can be found in our Github project repository~\cite{replication}.

Table~\ref{tab:claim_dataset} presents an overview of our $122$ misconceptions, spanning six categories: $(1)$~Crypto and Blockchain, $(2)$~IoT/CPS, $(3)$~Law and Regulation, $(4)$~Malware and Device Security, $(5)$~Privacy and Anonymity Tools, and $(6)$~Web Security and Privacy.

\subsection{LLM Response Generation}
\label{subsec:llm_response_generation}

\shortsectionBf{Model Selection.}
We focus on two prominent and widely-used LLM chatbots, ChatGPT and Bard, to evaluate LLM responses to these S\&P misconceptions. 
We selected them as $(a)$~they are the most popular LLM chatbots available today, $(b)$~they offer user-friendly interfaces accessible to the general public, and $(c)$ do not require fine-tuning before end-user interaction. 

We note that we did not evaluate S\&P misconceptions with chatbots, such as BlenderBot~\cite{shuster2022blenderbot}, DialogGPT \cite{zhang2019dialogpt}, and GODEL~\cite{peng2022godel}. 
These chatbots are not explicitly designed to provide comprehensive advice or engage in conversations on niche topics. % (\eg S\&P).
To enable them to answer such questions, one would require further fine-tuning with a vast corpus of relevant information.

In our preliminary experiments, we randomly sampled $50$ misconceptions from our dataset and queried BlenderBot, DialogGPT, and GODEL for their opinions on these misconceptions. 
They showed a lack of awareness for most S\&P questions. 
To illustrate, when we asked about their opinion on the misconception; 
\begin{quote}
\textit{``Pseudonymised data (\eg hashed data) are treated exactly like any other personal data under the GDPR''},  
\end{quote} 
they returned responses such as;
\begin{quote}
    \textit{``I do not have information to answer this''},  \textit{``I don't know, I don't work for the government''} or \textit{``yes, that's my guess''} 
\end{quote} 
Therefore, we excluded such language models from our analysis due to their inadequate performance.

\shortsectionBf{Response Generation Experiments.}
To generate responses from LLMs, we framed each misconception as a claim and queried the LLM to verify it. 
We used the following static template as the input prompt: \textit{I've heard of this claim: \{MISCONCEPTION\}. Is it true?}. 
We designed this template to mimic end users who interact with state-of-the-art LLM interfaces without access to more complex prompt templates (\eg fine-tuning prompts via prompt engineering~\cite{kojima2022large, wei2022chain}).
To maintain consistency, we use the most updated versions for both models at the time of writing this paper. 
We query ChatGPT using its official API~\cite{chatgptapi}, using the latest version (the gpt-3.5-turbo~\cite{gpt3.5}).
We follow the API-call examples provided on the official website and use the default parameters throughout our interactions.
For Bard, since we have no access to its official API, we query misconceptions (using our prompt template) via Bard's web interface and obtain responses. 
After each query, we refresh the chat and delete the activity history to prevent interference between queries.
We note that data collection for our experiments was conducted in a 2-day span ($14$, $15$ March 2023). 
We did so to minimize the impact of the potential updates introduced by Google and OpenAI for Bard and ChatGPT, respectively.
We designed four experiments (\texttt{E1-E4}) to extensively evaluate LLM capability in providing S\&P advice, focusing on their (1) correctness in refuting misconceptions, (2) consistency in providing the same stance towards a misconception, (3) susceptibility to different framings/paraphrases of misconceptions and (4) ability to provide reliable sources.

\shortsectionBfn{Initial Analysis via Single Trial (\texttt{E1}).}
In \texttt{E1}, we query each misconception one time (a single trial), for both ChatGPT and Bard.

\shortsectionBfn{Repeated Queries (\texttt{E2}).} 
In \texttt{E2}, we evaluate the consistency of LLMs in responding to S\&P misconceptions. 
The models that users interact with (via web interfaces) are non-deterministic - asking the same question twice may not result in identical responses~\cite{chatgpt-api-temperature}.
To simulate real-world scenarios where multiple individuals may ask the same question and receive different responses, we conducted four additional trials per misconception, generating a total of $488$ additional responses from each model.

\shortsectionBfn{Paraphrased Queries (\texttt{E3}).} 
In \texttt{E3}, we evaluate the effectiveness of LLMs in handling paraphrased queries since users may query LLMs chatbots in various ways. 
To do so, we use paraphrasing tools to produce an augmented dataset consisting of paraphrases of our original misconception. 
There is a wide selection of commercial and open-source paraphrasing tools; however, their performance in maintaining the original meaning of a sentence varies significantly. 

To identify the most suitable tools, we conduct preliminary analysis on $27$ APIs on the Rapid API platform~\cite{rapidapi} and $10$ open-source paraphrasing models on the Hugging Face platform~\cite{huggingface}, two resources that provide state-of-the-art AI tool APIs.
We randomly select five misconceptions from our dataset and generate paraphrases for each misconception with each tool.

\input{0-ACSAC-Minor-Revision/Content/tables/e3-paraphrases}

\input{0-ACSAC-Minor-Revision/Content/tables/label-examples}

Similar to determining manuscript relevance, two authors independently examined each paraphrased misconception using a labeling guide, labeling them as \textit{valid} if they remained coherent while expressing the same meaning as the original misconception and \textit{invalid} otherwise.
Authors achieved high agreement, $\kappa>0.80$ before differences were reconciled.
Based on the quality of the generated paraphrases, we chose Paraphrase Genius~\cite{paragenius} available via Rapid API and t5-large-paraphraser-diverse-high-quality model~\cite{t5-paraphrase} on Hugging Face, since these two models returned the highest percentage of \textit{valid} paraphrased misconceptions.

We generate as many paraphrases as possible for each misconception using these tools. 
Then for each misconception, we manually select four paraphrases of high quality, which fluently and accurately convey the complete meaning of the original misconception. 
Table~\ref{tab:e3-paraphrases} presents a set of examples that we generated through this process.
Overall, with the paraphrases, we generated $488$ additional responses from Bard and ChatGPT.

\shortsectionBf{Soliciting Sources (\texttt{E4}).}
We queried LLMs to obtain the URLs of their information to evaluate ability in providing reliable sources when responding to a misconception.
We followed up on queries in \texttt{E2} with the prompt \textit{``Can you provide the URLs of your source?''}.

\subsection{LLM Response Analysis}
After we collect the responses from each LLM, we label them into different categories according to their \emph{stances} on the misconception.

\shortsectionBf{Labeling Guide Development.}
To develop our labeling guide, we randomly sample $30$ responses from \texttt{E1} for both LLMs.
Two authors independently performed deductive coding on the generated $60$ responses,  coding each response while focusing on the stance towards the misconception (whether they confirm or deny the misconception and to what degree). 
This allows us to produce a labeling guide that characterizes a response into one of four categories:

\begin{itemize}
\setlength\itemsep{0.5em}
    \item \texttt{Noncommittal}: This category encompasses responses where the LLMs express a lack of knowledge about the topic or are unable to take a definitive stance. 
    \item \texttt{Negate}: Responses falling under this category highlight the presence of falsehood or inaccuracies in the misconception. 
    \item \texttt{Support}: Responses that affirm the validity or truthfulness of the misconception are classified under this category. 
    \item \texttt{Partially Support}: Responses that confirm the misconception's validity to a limited degree but do not address any shortcomings or falsehoods in the misconception. 
\end{itemize}

Table~\ref{tab:labels} shows an example misconception and its corresponding response for each label .
We introduce an additional label \texttt{Unrelated} to address outliers, where a response may deviate from the main question or fail to address specific information the misconception inquires about.
Given that our dataset comprises common S\&P misconceptions, the ground truth for each response is \texttt{Negate}.

\shortsectionBf{Guide Validation and Labeling Process.} To ensure our labeling guide is reliable, we randomly sample $10$ misconceptions and label their corresponding responses from \texttt{E2}, for a total of $100$ responses.
Two authors then independently labeled each response according to the labeling guide. 
We measure the agreement using Cohen's Kappa and obtain high agreement ($k\ge 0.80$). 
The authors met to reconcile differences and agreed on the final guide version.
They then labeled all remaining responses independently, meeting at intervals to reconcile differences.
We note that the initial agreement (before differences are discussed) is high ($k\ge 0.80$) at every interval.

\shortsectionBf{Correctness and Error Rate Analysis.}
We evaluate correctness on a per-claim (misconception) basis.
For \texttt{E1}, we consider responses labeled \texttt{Support} incorrect, as they indicate when an LLM fails to refute the misconception and provide incorrect support. 
In repeated trial experiments (\texttt{E2}-\texttt{E3}), we adopt a conservative approach and consider the misconception result incorrect if \textit{any} of the trials result in a response labeled \texttt{Support}. 
Similarly, we consider responses labeled \texttt{Negate} correct, and for repeated trial experiments, a misconception producing accurate results must have \textit{Negate} responses across \textit{all} trials.
We defined \emph{error rate} as the ratio of misconceptions producing an incorrect result over total number of misconceptions.

\shortsectionBf{Analyzing Source URLs.}
To analyze collected URLs, we first leverage the Python HTTP library \texttt{requests}~\cite{requests} to verify if the website exists at present.
If the request is successful (no error code is returned), we label it as a valid URL.
If we receive an error, we employ an additional check using the Wayback Machine API~\cite{wayback_machine} to verify if the directed website once existed.
If it had existed but has since been removed (\eg expired domain), we label it as a valid URL.
If the URL is invalid (\texttt{requests} returns an error, and the Wayback Machine has no archive of the URL),  we label it as an invalid URL.

We further label all valid URLs, as one of three categories, in accordance with their domain relevance.
We label the URL $(1)$~\textit{relevant} if it directs to a website that provides S\&P advice/information that relates to the misconception's domain (\eg misconception on VPN resulting in a website about VPN S\&P).
We label it $(2)$~\textit{marginally relevant} if the directed website provides S\&P advice that is generic or unrelated to the misconception, \eg the misconception is about password security, but the website outlines VPNs security and privacy capabilities with no mention of password security. 
Lastly, we label a URL $(3)$~\textit{irrelevant} if the website content is unrelated to S\&P (\eg a website on iPhone features that does not mention S\&P).

%% file: 0-ACSAC-Minor-Revision/Content/tables/dataset-table.tex
\begin{table*}[th!]
\centering
\caption{Overview of our S\&P misconception dataset, which is categorized into one of six categories.}
\label{tab:claim_dataset}
\setlength{\tabcolsep}{1.3em}
\def\arraystretch{1.2}
\resizebox{\linewidth}{!}{
\begin{tabular}{p{4cm}||P{2.5cm}|p{14cm}}
\hline
\textbf{Category} & \textbf{$\#$ Misconceptions} & \textbf{Example Misconceptions}
\\\hline\hline
Crypto and Blockchain   & 25 & Every transaction on the blockchain is anonymous.
\\\hline
IoT/CPS	    &17 & In smart homes, only devices I actively interact with are able to collect data about me (e.g., the doorbell).
\\\hline
Law and Regulation  &14     & Under GDPR, when relying on consent to process personal data, consent must be explicit.
\\\hline
Malware and Device Security	&21 & My PC or network cannot be harmed by my visiting a website, if I don't download anything.
\\\hline
Privacy and Anonymity Tools	&25 & Employers would be unable to track employees when they used private mode.
\\\hline
Web Security and Privacy    &20 & Websites that use HTTPS are trustworthy.
\\\hline
\end{tabular}
}
\end{table*}

%% file: 0-ACSAC-Minor-Revision/Content/tables/e3-paraphrases.tex
\begin{table*}[th!]
\centering
\caption{Example of generated paraphrases for misconceptions.}
\label{tab:e3-paraphrases}
\setlength{\tabcolsep}{1.2em}
\def\arraystretch{1.1}
\resizebox{\linewidth}{!}{
\begin{tabular}{l|p{16cm}}
\hline
\textbf{Misconception} & \textbf{Paraphrases} \\\hline\hline
\multirow{4}{5cm}{Pseudonymised data (e.g., hashed data) are treated exactly like any other personal data under the GDPR.} 
& According to the GDPR, pseudonymized data (such as hashed data) is treated in the same way as other personal data. \\ \cline{2-2} 
& Pseudonymized records (\eg hashed records) are treated in the same way as any other personal information under GDPR. \\ \cline{2-2} 
& The GDPR treats pseudonymized data (such as hashed data) the same as other personal data. \\ \cline{2-2} 
& Under the GDPR, pseudonymized data (such as hashed data) are treated in the same way as other personal data. \\ \hline
\multirow{4}{5cm}{If I do not open an infected file, it can't do any harm.} 
& A malicious file cannot cause me harm if I don't open it. \\ \cline{2-2} 
& I won't be harmed if I don't open a malicious file. \\ \cline{2-2} 
& An infected file can't harm me if I don't open it. \\ \cline{2-2} 
& If I do not open an infected file, it won't do me any harm. \\ \hline
\multirow{4}{5cm}{VPNs would prevent hackers from gaining access to their device.} 
& Hackers wouldn't be able to access their devices thanks to VPNs. \\ \cline{2-2} 
& Using a VPN would stop hackers from accessing their devices. \\ \cline{2-2} 
& Virtual private networks (VPNs) would stop hackers from accessing their devices. \\ \cline{2-2} 
& Through the use of VPNs, hackers could not access the user's device. \\ \hline 
\end{tabular}
}
\end{table*}

%% file: 0-ACSAC-Minor-Revision/Content/tables/label-examples.tex
\begin{table*}[ht!]
\centering
\caption{Labels and example misconceptions and responses. Relevant text that informs the labeling decision is highlighted.}
\label{tab:labels}
\setlength{\tabcolsep}{1em}
\def\arraystretch{1.1}
\resizebox{\linewidth}{!}{
\begin{tabular}{p{2.5cm}|p{16cm}}
\hline
\textbf{Label} & \textbf{Example claim and corresponding response.}
\\\hline \hline
\multirow{2}{*}{\texttt{Noncommittal}} 
& \textbf{Claim}: VPNs would be not at all effective at preventing observation by the ISP.\\ \cline{2-2} 
&  \textbf{Response}: As an AI language model, \hl{\mbox{I don’t have personal opinions or beliefs}}, but I can  provide information on the topic. It is not entirely true or false. 
\\ \hline
\multirow{2}{*}{\texttt{Negate}}
& \textbf{Claim}: Under GDPR, individuals have an absolute right to be forgotten. \\ \cline{2-2} 
& \textbf{Response}: \hl{No, it is not true.} The General Data Protection Regulation (GDPR) gives individuals the right to request that their personal data be erased, but this right is not absolute. \\ \hline
\multirow{2}{*}{\texttt{Support}}
& \textbf{Claim}: Malware is most commonly downloaded through peer-to-peer and torrent sites.\\ \cline{2-2} 
& \textbf{Response}: \hl{Yes, it is true} that malware is most commonly downloaded through peer-to-peer and torrent sites.  \\ \hline
\multirow{2}{*}{\texttt{Partially Support}}
& \textbf{Claim}: Tor browser is for the dark web and file sharing; it is the very definition of a dark web browser.\\ \cline{2-2} 
& \textbf{Response}: As an AI language model, I cannot say whether it is true or false as \hl{it can be subjective. However, it is true that Tor is often used to access the dark web}, which is a part of the internet that is not indexed by standard search engines and requires specific software to access it.\\ \hline
\end{tabular}
}
\end{table*}

%% file: 0-ACSAC-Minor-Revision/Content/evaluation.tex
\section{Evaluation}
\label{sec:eval}
In this section, we detail our findings from our experiments \texttt{E1-E4}.

%%%%%%%%TABLE INPUT%%%%%%%%
\input{0-ACSAC-Minor-Revision/Content/tables/e1-response-stats}

%%%%%%%%TABLE INPUT%%%%%%%%

%
% 
\subsection{Single Trial Queries (E1)}
\textbf{Correctness.} In the first experiment, we ask ChatGPT and Bard each misconception once. Table~\ref{tab:exp1-ask-one-time} presents the distribution of responses in \texttt{E1}. 
We find that Bard correctly negates $72.1$\% of the misconceptions. And it has an error rate of $26.2$\%, where it incorrectly supports the misconceptions. 
As for ChatGPT, we find that it correctly negates $70.5$\% of the claims but has an error rate of $16.4$\%.
It is important to note that ChatGPT provides a \textit{``noncommittal''} response in $9.84\%$ of trials, compared to only $1.64\%$ of trials in Bard. 
These empirical results demonstrate that Bard is less likely to refuse to answer a claim/remain neutral.

\begin{figure}[t!]
    \centering
\includegraphics[width=\columnwidth]{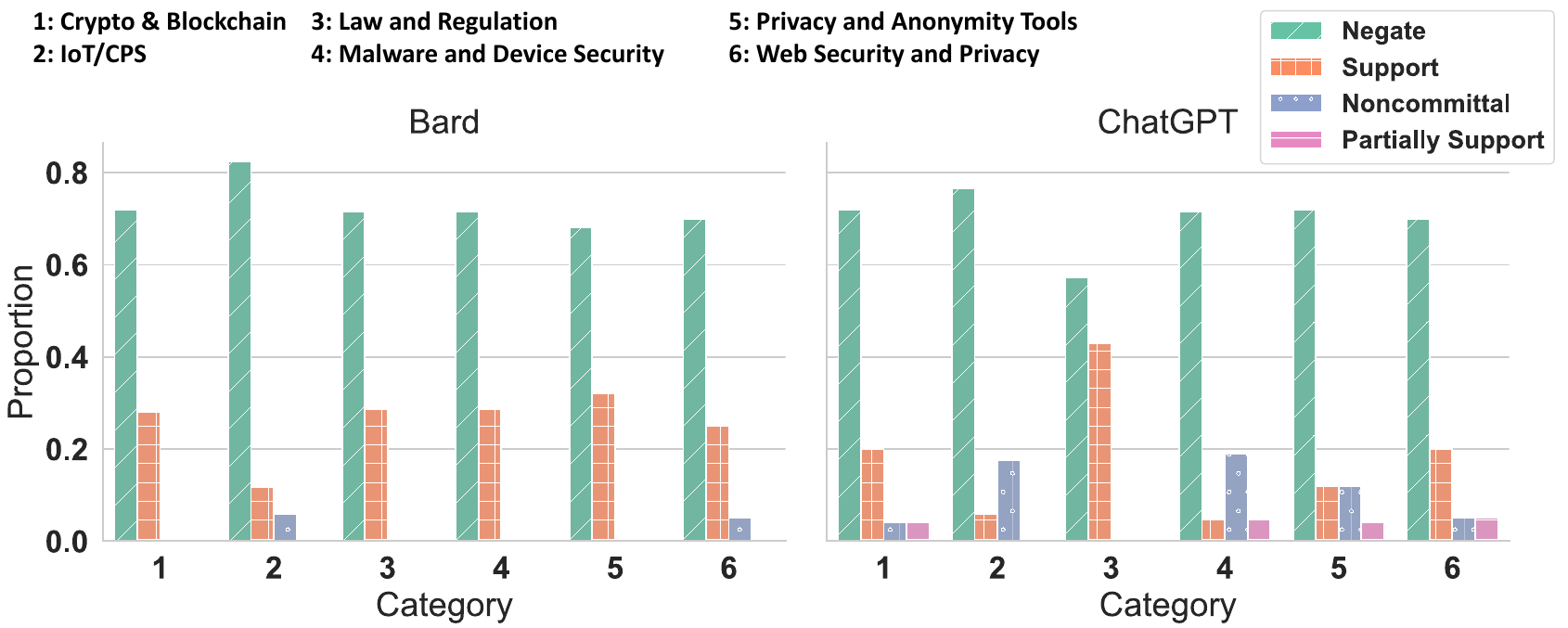}
    \caption{Proportion of each response label across misconception categories  (\texttt{E1}).} 
    \label{fig:e1-res-label-per-category-normalized-count}
\end{figure}

\shortsectionBf{Comparing Response Categories.} We performed additional analyses to study ChatGPT and Bard's performance across different categories. 
Figure~\ref{fig:e1-res-label-per-category-normalized-count} shows the proportions of each response label in ChatGPT and Bard's responses to these misconceptions. 
These proportions are calculated with respect to the number of misconceptions in each category. 
We interpret the \textit{Negate} proportion as the correct rate and the \textit{Support} proportion as the error rate. 
For instance, we find that ChatGPT exhibits the highest error rate for misconceptions falling under the ``Law and Regulation'' category ($42.9\%$), while misconceptions pertaining to ``Malware and Device Security'' had the lowest error rate ($4.76\%$), albeit the highest rate of noncommittal responses ($19.1\%$). 
For Bard, ``Privacy and Anonymity Tools''  produces the highest error rate ($32\%$).
For the rest of the categories, error rates are around $25\%$, except for ``IoT/CPS'', which demonstrates an $11.8$\% error rate. 

\begin{shadedboxed}
Our empirical results in \texttt{E1} highlight that although both models correctly \textit{Negate} misconceptions \textasciitilde~$70\%$ of the time, they also demonstrate a non-negligible error rate.
\end{shadedboxed}

%
 
%%%%%%%%%TABLE INPUT%%%%%%%%
\input{0-ACSAC-Minor-Revision/Content/tables/e2-unique-labels}

%%%%%%%%TABLE INPUT%%%%%%%%

\subsection{Effectiveness under Repeated Queries (E2)}
\label{subsec:e2}

\begin{figure}[t!]
    \centering
\includegraphics[width=\columnwidth]{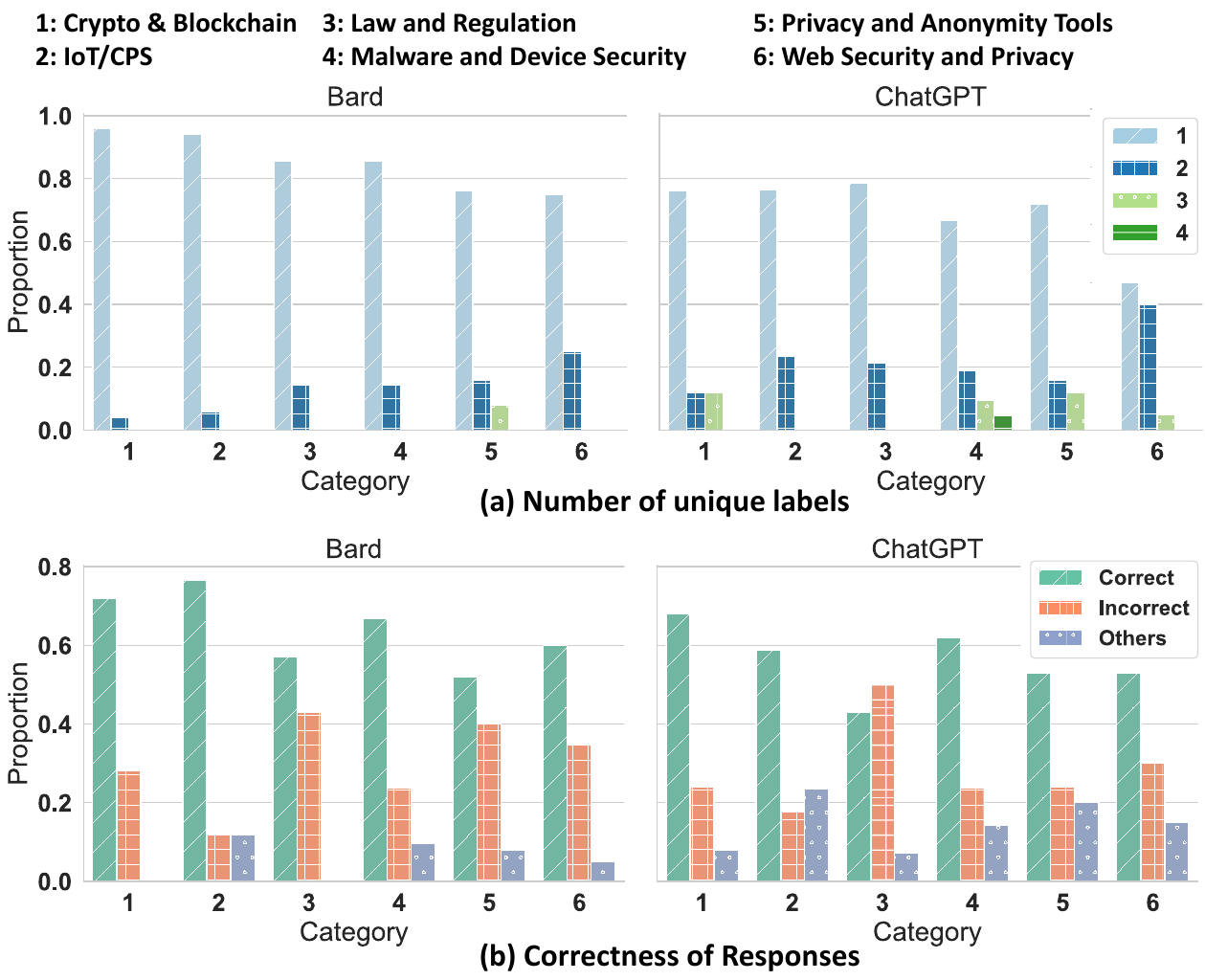}
    \caption{Results across misconception category for E2.} 
    \label{fig:e2}
\end{figure}

\shortsectionBf{Consistency of Responses.} 
In this experiment, we queried each misconception four additional times for ChatGPT and Bard, making a total of 610 responses for each model when combined with \texttt{E1}.
Table~\ref{tab:exp2-ask-five-times-unique-labels} presents the number of unique response labels for each misconception. 
Bard performs better than ChatGPT in remaining consistent in their stance toward a misconception.
$85.2$\% of Bard's misconceptions produce single label responses, compared to ChatGPT's  $70.5$\% 
of misconceptions.
However, both Bard and ChatGPT demonstrate a non-negligible tendency to be inconsistent toward misconceptions  ($14.8$\% and $29.5$\% of misconceptions, respectively, produce more than one type of response across all trials).

We further examined consistency across misconceptions categories, as shown in Figure~\ref{fig:e2}~(a).
For Bard, inconsistency typically remains relatively low across all categories (below the $25\%$ threshold). 
Noticeably, for both Bard and ChatGPT, ``Web Security and Privacy'' and ``Privacy and Anonymity Tools'' are the categories that they are most prone to changing stances to ($25$\% and $24$\% for Bard, and $45$\% and $28$\% for ChatGPT).
%

%%%%%%%%TABLE INPUT%%%%%%%%
\input{0-ACSAC-Minor-Revision/Content/tables/e2-label-stats}
%%%%%%%%TABLE INPUT%%%%%%%%

\shortsectionBf{Influence of Repeated Queries on Correctness.}
In Table~\ref{tab:exp2-ask-five-times-res}, we present the response distributions.
We observe an increase in error rate as a result of repeated queries. 
More specifically, Bard and ChatGPT show an error rate of $30.3$\% and $27$\%, respectively. 
This translates to $4.1$\% and $10.6$\% increase from the results of \texttt{E1}.
Bard shows a higher correct rate than ChatGPT, which can be attributed to Bard's tendency to maintain consistency.
As expressed in Table~\ref{tab:exp2-ask-five-times-unique-labels}, Bard is $14.7\%$ more likely to maintain a single stance than ChatGPT.
Here, we note that ChatGPT's tendency to produce more than one response type causes a drop in its correctness.

We note that if responses are neither correct (all trials \textit{Negate} the misconception) nor incorrect (any \textit{Support} across trials), we group them under ``Others''.
For example, for the misconception \textit{``Tor browser is for the dark web and file sharing, it is the very definition of a dark web browser.''}, we find that Bard negates four times, and is noncommittal once. 
Such scenarios highlight that although an LLM may never incorrectly support a misconception, they can still be unreliable.
Despite having previously negated a misconception, they are susceptible to taking a less concrete stance (\textit{noncommittal}).
Depending on randomness, this can disadvantage users, who may receive a noncommittal response towards a misconception (that should be negated). 
Thus, they do not receive the required S\&P advice.
$5.8$\% and  $14.8\%$ of Bard and ChatGPT responses towards misconceptions are grouped as ``Others''.

Figure~\ref{fig:e2}~(b) shows the proportions of each response type across misconception categories.
Among the categories, ``Law and Regulation'' exhibits the highest error rates in ChatGPT and Bard, $50$\% and $42.9$\%. 
We posit that law and regulation produce higher error rates as the language used may require more context. 
To illustrate, users may query LLMs with the misconception \textit{``Under GDPR, when relying on consent to process personal data, consent must be explicit''} while understanding how ``explicit'' is used in the context of privacy laws. 
Surprisingly, our findings show that LLMs misinterpret ``explicit'' and confuse it with ``unambiguous'', which bears different implications in a legal context. %

On the other hand, the ``IoT/CPS'' and ``Malware and Device Security'' misconception categories have the lowest error rates. 
For IoT/CPS, ChatGPT and Bard have error rates of $17.7$\% and  $11.8$\%, respectively.
Both ChatGPT and Bard demonstrate an error rate of $23.8$\% for ``Malware and Device Security''.

\shortsectionBf{Confusing Responses.} 
We found that a minority of responses contain confusing patterns - where the response begins with language indicating agreement for the misconception but provides context negating it.
$2.95$\% and $0.984$\% of responses for ChatGPT and Bard begin with \emph{Yes, it is true}, but the remainder of the text \emph{negates} the claim.
Such responses are labeled with \textit{Negate} as we consider the context of the entire response in our labeling process.
For instance, when asked about the misconception \textit{``Under GDPR, individuals have an \textbf{absolute} right to be forgotten.''}, ChatGPT responds with 
\begin{quote}
    \textit{``\textbf{Yes, it is true} that under GDPR... individuals have a ``right to be forgotten'' ... 
    However, \textbf{this right is not absolute} and there are limitations and exemptions. For example, organizations may be permitted to retain certain data for legal or regulatory reasons.''}
\end{quote}

\noindent Since ChatGPT and Bard's responses tend to be elaborate,  users who do not pay great attention may be misled by these responses.

\begin{shadedboxed}
Both models show a non-negligible tendency to be inconsistent in their stance when they are queried with the same misconception; however, they also yield an increased error rate. 
Responses from both models also contain confusing patterns that may mislead unassuming users. 
\end{shadedboxed}

%%%%%%%%TABLE INPUT%%%%%%%%

%%%%%%%%TABLE INPUT%%%%%%%%

\subsection{Effectiveness on Paraphrased Queries (E3)} 
\label{subsec:e3}

\begin{figure}[t!]
    \centering
\includegraphics[width=\columnwidth]{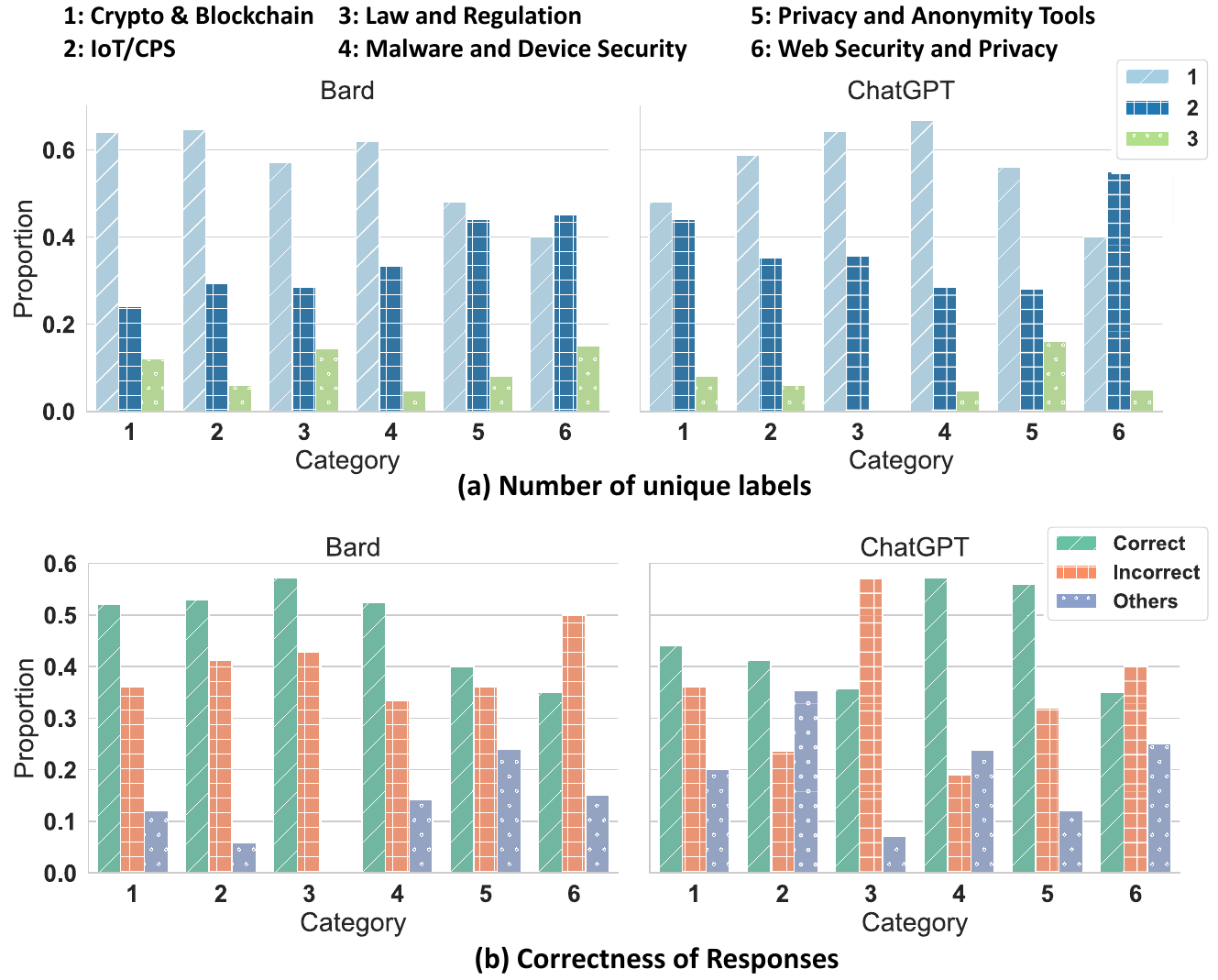}
    \caption{Results across misconception category for E3.} 
    \label{fig:e3}
\end{figure}
\shortsectionBf{Consistency of Responses.} 
In this experiment, we query the four paraphrases of each misconception once on each model and analyze these responses together with those in \texttt{E1}.
We discover a significant increase (when compared to \texttt{E2}) in inconsistency (misconceptions with two or more label types).  

\input{0-ACSAC-Minor-Revision/Content/tables/e3-unique-labels}
Table~\ref{tab:exp3-paraphrases-unique-label} presents the distribution of unique labels across misconceptions.
We find that, on average, $44.7$\% of misconceptions in Bard and ChatGPT solicit inconsistent responses. 
This is an increase of $29.5$\% and $15.6$\% in Bard and ChatGPT, when compared to \texttt{E2}, suggesting that slight modifications (while maintaining the same meaning) to sentences significantly decrease LLMs' consistency.

To demonstrate this, in Figure~\ref{fig:e3}~(a), we show the distribution of unique label types per category.
For both ChatGPT and Bard, the category ``Web Security and Privacy'' exhibits the highest occurrence of inconsistency, $60$\% for both models. 
%

%%%%%%%%TABLE INPUT%%%%%%%%
\input{0-ACSAC-Minor-Revision/Content/tables/e3-label-stats}
%%%%%%%%TABLE INPUT%%%%%%%%

\shortsectionBf{Response Correctness.} Table~\ref{tab:exp3-paraphrases-res} presents an overview of correctness in \texttt{E3}.
Unsurprisingly, following an increase in inconsistency, overall correctness decreases by $16.4$\% and $12.3$\% for Bard and ChatGPT, respectively (compared to \texttt{E2}). 
Compensating for this drop, we observed a higher error rate in Bard and ChatGPT (an increase of $9$\% and $6.6$\%, respectively, from \texttt{E2}) and in ``Other'' responses (an average increase of $6.55$\%). 
Interestingly, Bard's correct rate is only slightly higher than ChatGPT's (by $1.6\%$),  contrasting \texttt{E2}'s results.  
We attribute the similarity in correctness to both models' shared vulnerability to paraphrased questions - they both stick to the same stance across all questions only around $55\%$ of the time.
Figure~\ref{fig:e3}~(b) presents correctness across misconception categories.
We observed that the categories ``Law and Regulation'' and ``Web Security and Privacy'' continue to exhibit the highest error rates.

\begin{shadedboxed}
    Paraphrasing queries reduces LLM consistency and causes an significant increase in error rate and reduction in correctness when compared to multiple questions for a misconception.
\end{shadedboxed}

\input{0-ACSAC-Minor-Revision/Content/tables/source-analysis-table}
\subsection{Analysis of Response Sources (E4) }
\label{subsec:e4}

\begin{figure}[t!]
    \centering
\includegraphics[width=1.01\columnwidth]{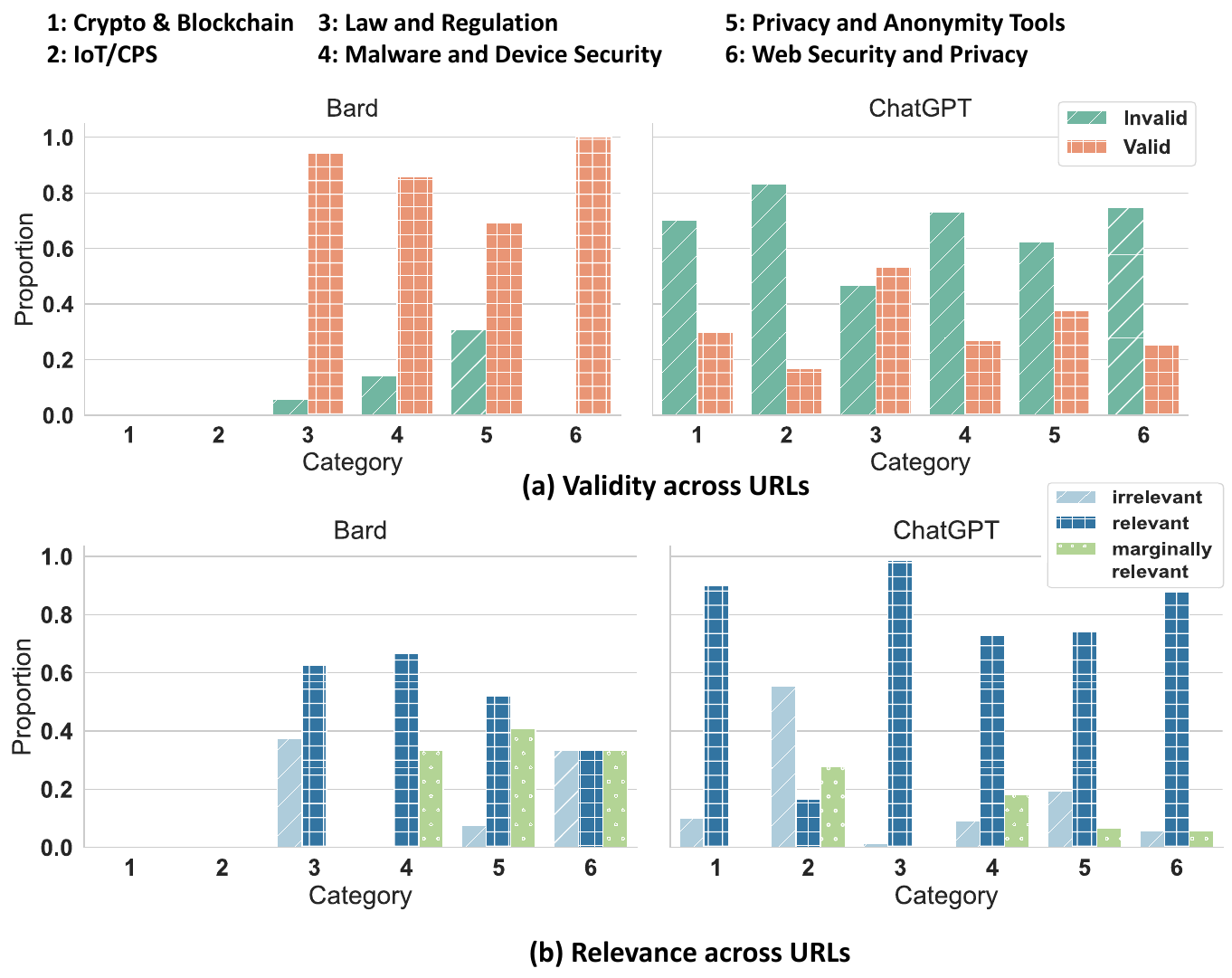}
    \caption{Results across misconception category for E4.} 
    \label{fig:e4}
\end{figure}
\shortsectionBf{URL Validity and Relevance.} 
We gathered $946$ URLs from ChatGPT's responses obtained in \texttt{E2} from $108$ misconceptions, with an average of $8.76$ URLs per misconception.
The remaining $14$ misconceptions do not produce any URLs. 
To illustrate, when asked for sources on the misconception \textit{``Tor is designed for criminals who want to do illegal business securely.''}, ChatGPT responds
with 
\begin{quote}
    \textit{``As an AI language model, I don't have the ability to browse the internet and provide URLs of sources.''}
\end{quote}

Interestingly, we find that Bard does not include explicit URLs in the responses.
None of the responses to our prompt in the form of \textit{Can you provide the URL of your sources?} provide URLs, with Bard always refusing (\eg \textit{``
I'm just a language model, so I can't help you with that''}).
We are only able to collect $66$ URLs from the ``Source'' section, which occasionally comes together with the response to the question on misconception. 
% Bard occasionally responds with sources in their response to a misconception -  we collected $66$ URLs from Bard's responses. 
%
These URLs are spread across $29$ misconceptions, with an average of $2.28$ URLs per misconception.
We find that $78.8$\% of URLs returned by Bard are valid - $71.23$\% exist currently, $7.57$\% no longer exist but can be accessed via the internet archive~\cite{wayback_machine}.
Conversely, only about a third ($32.3$\%) of URLs returned by ChatGPT are deemed valid, $27.1$\% exist currently, and $5.2$\% no longer exist but are archived. 
This stark difference between both models is expected given that Bard has real-time access to the internet, while ChatGPT does not. 
However, given Bard's real-time access, it is still surprising that $21.2$\% of Bard's URLs are invalid. 
Notably, for Bard,  $9.62$\% of the valid URLs were currently inaccessible because of SSL connection errors. 
Figure~\ref{fig:e4}~(a) presents URL validity across the misconception categories. 
It's worth noting that Bard provides no URLs for the ``Crypto \& Blockchain'' and ``IoT/CPS'' misconception categories. 
Furthermore, Bard consistently provides a higher percentage of valid URLs across all categories. 
In contrast, ChatGPT provides more invalid URLs across all categories.

We additionally explored the domain relevance for each website the valid URL directs to.  
Figure~\ref{fig:e4}~(b) shows URL domain relevance across the misconception categories. 
For Bard, only $55.8$\% were \textit{relevant}, $26.9$\%  were \textit{marginally relevant} - unrelated to the domain but still contained security and privacy content, and the remaining $17.3$\% were entirely unrelated to the claim or its response.
For ChatGPT, $81.7$\% of domains were \textit{relevant}, $5.9$\% of them were \textit{marginally relevant}, and the rest ($12.4$\%) had no connection to the claims, \textit{irrelevant}. 
Fortunately, the relevant URLs comprise the majority of valid URLs across all categories. 
However, $55.6$\% of valid URLs returned for the ``IoT/CPS'' category in ChatGPT are \textit{irrelevant} (the highest percentage of invalid across all categories).

\shortsectionBf{Understanding Sources that Result in Errors.} We further analyze the URLs provided in incorrect responses. 
For Bard, $38.5$\% of returned valid URLs are attributed to incorrect responses, while this percentage for ChatGPT is $35.6$\%.
To gain further insight into the LLMs' error generation, we read websites the valid URLs direct to.
Given Bard's relatively small URL pool satisfying this category ($20$ URLs), we analyzed all of the websites that it provides. 
%associated
For ChatGPT, we sampled one-third of the valid URLs with incorrect responses. 

Table~\ref{tab:exp4-source-analysis-bard} shows example misconceptions with incorrect responses, their URLs, and corresponding analysis.
In \circled{1}, the website ignores cases where parental consent is not necessary when processing children's personal data (\textbf{false information}).
Similarly, in \circled{2}, the website failed to point out that under certain circumstances the blockchain ledger could be possibly changed. 
In \circled{3}, the website provides \textbf{generic information} on VPNs' encryption capabilities without addressing the privacy features relevant to the misconception (the ability of friends/family to see websites in browser history). 
\circled{4}  shows Bard's failure to comprehend an official GDPR article fully, thus leading to an incorrect response (\textbf{partial information}).
In \circled{5}, it is clearly stated in the website that new passwords and bookmarks created while using private browsing will be saved, while ChatGPT ignores it (\textbf{partial information}).
In \circled{6}, the website has no relation to the claim about the government's ability to access browsing activity in private mode.

Upon inspection of corresponding web pages associated with incorrect responses, $70$\% of the valid URLs Bard provided were domain \textit{relevant}. 
Comprising this, $35$\% sources only provided generic information, $10$\% provided false information. $25$\% provided correct information, but Bard responded with partial information resulting in an incorrect response. 
The remaining $30$\% were \textit{``marginally relevant''}/\textit{``irrelevant''} - they do not address the misconception domain.

For ChatGPT, 54.8\% sources offered generic information without any direct statements on the misconception, and  9.5\% contained false information. 
23.8\% URLs contain the correct information, but, ChatGPT responds with partial information. 
The remaining 11.9\% URLs in the sample were domain  \textit{``marginally relevant''}/\textit{``irrelevant''}.

\begin{shadedboxed}
    Bard divulges sources less frequently than ChatGPT. However, Bard's URL sources are more likely to be valid.  
    Valid URLs predominantly point to websites that are domain \textit{relevant} to the misconception category, but may also be unrelated to the misconception. 
    For incorrect responses that provide relevant URLs, websites predominantly only have generic information on the misconception domain.  
    %
    %
    %
%Arjun will write this after Yufan fills in blanks.
\end{shadedboxed}

%% file: 0-ACSAC-Minor-Revision/Content/tables/e1-response-stats.tex
\begin{table}[h!]
\centering
\caption{Response label distribution in single query (\texttt{E1}).}
\label{tab:exp1-ask-one-time}

\resizebox{\columnwidth}{!}{
\begin{threeparttable}
\begin{tabular}{c|c|c|c|c|}
\cline{2-5}
& \textbf{Negate$^\dagger$} & \textbf{Support$^\ddagger$} & \textbf{Partially Support} & \textbf{Noncommittal}   \\\hline
\multicolumn{1}{|l|} {\textbf{Bard}} & 72.1\% & 26.2\% & 0\% & 1.7\%\\\hline
\multicolumn{1}{|l|} {\textbf{ChatGPT}} & 70.5\% & 16.4\% & 3.3\% & 9.8\%\\\hline
\end{tabular}
    \begin{tablenotes}
          \small \item $^\dagger$ represents claims correctly responded to and $^\ddagger$ represents the error rate. 
        \end{tablenotes}
\end{threeparttable}
}
\end{table}

%% file: 0-ACSAC-Minor-Revision/Content/tables/e2-unique-labels.tex
\begin{table}[ht!]
\setlength{\tabcolsep}{1em}
\def\arraystretch{1}
\centering
\caption{Distribution of unique labels in repeated queries (\texttt{E2}). More than one unique label demonstrates the LLM's inconsistency towards a misconception.}\label{tab:exp2-ask-five-times-unique-labels}
\resizebox{\columnwidth}{!}{
\begin{threeparttable}
\begin{tabular}{c|P{1.2cm}|P{1.2cm}|P{1.2cm}|P{1.2cm}|}
\cline{2-5} & \textbf{1$^\dagger$ } & \textbf{2$^\dagger$} & \textbf{3$^\dagger$ } & \textbf{4$^\dagger$ } \\\hline 
\multicolumn{1}{|l|}{\textbf{Bard}} & 85.2\%  & 13.1\% & 1.7\% & 0\%\\\hline
\multicolumn{1}{|l|}{\textbf{ChatGPT}} & 70.5\%  & 21.3\% & 7.38\% &0.82\%\\\hline
\end{tabular}
  \begin{tablenotes}
          \small \item $^\dagger$ represents N unique labels while $\%$ values are with respect to $122$ misconceptions (\eg Bard produces 2 unique labels for $13.1\%$ of misconceptions).
        \end{tablenotes}
\end{threeparttable}
}
\end{table}

%% file: 0-ACSAC-Minor-Revision/Content/tables/e2-label-stats.tex
\begin{table}[h!]
\centering
\caption{Distribution of responses in repeated queries (\texttt{E2}).}
\label{tab:exp2-ask-five-times-res}
\setlength{\tabcolsep}{1.3em}
\def\arraystretch{1}
\resizebox{\columnwidth}{!}{
\begin{threeparttable}
\begin{tabular}{c|P{1.5cm}|P{1.5cm}|P{1.5cm}|}
\cline{2-4}
 & \textbf{Correct$^\dagger$} & \textbf{Incorrect$^\ddagger$} & \textbf{Others} \\\hline
\multicolumn{1}{|l|}{\textbf{Bard}} & 63.9\%  & 30.3\% & 5.8\% \\\hline
\multicolumn{1}{|l|}{\textbf{ChatGPT}} & 58.2\%  & 27.0\% & 14.8\%\\\hline
\end{tabular}
    \begin{tablenotes}
          {\small{\item $^\dagger$ represents claims correctly responded to. $^\ddagger$ represents the error rate.}} 
        \end{tablenotes}
\end{threeparttable}

}
\end{table}

%% file: 0-ACSAC-Minor-Revision/Content/tables/e3-unique-labels.tex
\begin{table}[h]
\centering
\caption{Distribution of unique labels in paraphrased queries~(\texttt{E3}).}
\label{tab:exp3-paraphrases-unique-label}
\setlength{\tabcolsep}{1.25em}
\def\arraystretch{1}
\resizebox{\columnwidth}{!}{
\begin{threeparttable}
\begin{tabular}{c|P{1.5cm}|P{1.5cm}|P{1.5cm}|}
\cline{2-4} & \textbf{1$^\dagger$} & \textbf{2$^\dagger$} & \textbf{3$^\dagger$} \\\hline
\multicolumn{1}{|l|}{\textbf{Bard}} & 55.7\%  & 34.4\% & 9.9\% \\\hline
\multicolumn{1}{|l|}{\textbf{ChatGPT}} & 54.9\%  & 37.7\% & 7.4\% \\\hline
\end{tabular}
  \begin{tablenotes}
          \small \item $^\dagger$ represents N unique labels while $\%$ values are with respect to $122$ misconceptions (\eg Bard produces 2 unique labels for $34.4\%$ of misconceptions).
        \end{tablenotes}
\end{threeparttable}
}

\end{table}

%% file: 0-ACSAC-Minor-Revision/Content/tables/e3-label-stats.tex
\begin{table}[h!]
\centering
\setlength{\tabcolsep}{1.1em}
\def\arraystretch{1}
\caption{Response distribution in paraphrased queries (\texttt{E3}).}
\label{tab:exp3-paraphrases-res}
\resizebox{\columnwidth}{!}{
\begin{tabular}{c|P{1.5cm}|P{1.5cm}|P{1.5cm}|}

\cline{2-4} & \textbf{Correct} & \textbf{Incorrect} & \textbf{Others} \\\hline
\multicolumn{1}{|l|}{\textbf{Bard}} & 47.5\%  & 39.3\% & 13.2\% \\\hline
\multicolumn{1}{|l|}{\textbf{ChatGPT}} & 45.9\%  & 33.6\% & 20.5\%\\\hline
\end{tabular}
}
\end{table}

%% file: 0-ACSAC-Minor-Revision/Content/tables/source-analysis-table.tex
\begin{table*}[t!]
\centering
\caption{Evaluating URLs provided with incorrect responses.}
\label{tab:exp4-source-analysis-bard}
\def\arraystretch{1.1}
\resizebox{\textwidth}{!}{
\begin{tabular}{p{8.5cm}|P{4.8cm}|p{6.5cm}}
\hline\textbf{Misconception} & \textbf{URL provided in LLM response} & \textbf{Evaluation} \\\hline\hline
\circled{1} Under GDPR, parental consent is always required when collecting personal data from children.
& \url{www.termsfeed.com/blog/childrens-gaming-apps-legal-requirements/} 
& False information (domain \textit{relevant}) \\\hline
\circled{2} The blockchain ledger is locked and unchangeable/ unable to modify the data block once created, or blockchain data cannot be changed once updated.
& \url{https://www.ibm.com/blockchain/what-is-blockchain} 
& False information (domain \textit{relevant}) \\\hline
\circled{3} VPNs would be very effective at preventing friends or family from seeing the websites in my browser history from my computer because I have my own private network that others cannot get into. 
& \url{https://quiaustin.com/do-online-casinos-track-your-ip-address-and-why/} 
& Generic information (domain \textit{relevant})
\\\hline
\circled{4} Under GDPR, every business will be subject to new data portability rules.
& \url{https://gdpr-info.eu/art-20-gdpr/} 
& Respond with partial information (domain \textit{relevant})
\\\hline
\circled{5} Bookmarks saved in private mode would not persist in later sessions because private mode deletes all local, temporary data, including bookmarks. 
& \url{https://support.mozilla.org/en-US/kb/private-browsing-use-firefox-without-history}
& Respond with partial information (domain \textit{relevant}) \\\hline
\circled{6} The government would need a warrant to access browsing activity from private mode. 
& \url{https://www.fights4rights.com/immigration-rights/} 
& Unrelated (\textit{irrelevant}) \\\hline
\end{tabular}
}
\end{table*}

%% file: 0-ACSAC-Minor-Revision/Content/discussion.tex
\section{Discussion and Limitations}
\subsection{Lessons and Recommendations}
\label{subsec:lessons}
\shortsectionBf{Shortcomings of LLMs in Providing Expert Advice.}
Our experiments show the limitations of LLMs in correctness, consistency, and susceptibility to paraphrasing.
We also expose their inability to justify or provide sources. 
These findings highlight how existing LLMs, in their current state, are unreliable as an S\&P advice tool.
LLMs are trained on vast amounts of web data. 
For instance, Wikipedia is among the most common web pages available in one dataset used to train Bard~\cite{raffel2020exploring, common_crawl_stats}. 
These resources are not guaranteed to be accurate (\eg Wikipedia edits can be approved by community members).
Additionally, as prior work has pointed out ~\cite{barrett2023identifying}, relying on web-scrapped data creates a vulnerable ``feedback loop'' - as inaccurate information generated by  AI models is often uploaded to the internet (\eg news articles synthesized by AI models~\cite{ali2019artificial}), influencing training data of current language models.

Similarly, despite numerous efforts to fine-tune LLMs, it remains unclear if domain-specific information (such as S\&P) is used in fine-tuning their state-of-the-art interfaces. 
Moreover, data labeling to fine-tune is often contracted to third parties instead of domain-specific experts (\eg researchers)~\cite{ouyang2022training}. 
We posit that these factors significantly contribute to LLM's lack of correctness. 
This is further evidenced in Section~\ref{subsec:e4}, by URLs pointing to generic and factually incorrect content, highlighting (1)~a lack of specialized and (2)~flawed training data, respectively. 

LLMs' growth in serving as a trusted resource, however,  shows no sign of slowing down. 
Soon after the introduction of LLMs,  LLM-based applications, and plugins emerged, claiming additional capabilities (\eg providing nutrition advice and identifying scientific sources)~\cite{gptstore, chatgpt-plugins}. 
Yet, the reliability of these tools in providing domain-specific expert advice has not been evaluated.
Future work should not only address the feasibility of training LLMs to provide expert advice to users but should require careful collaboration with domain experts (\eg health experts should be consulted during the design of LLM-driven health advice tools).

\shortsectionBf{Broadening Experimental Scope.}
Although our efforts provide insight into LLM's limitations in providing S\&P advice, future work should expand our data and experimental procedures. 
First, our dataset can be expanded in quantity and category diversity. 
This can be achieved via community collaboration (\eg with S\&P experts in various domains). 
Second, through larger-scale data collection and labeling, future efforts should also train an ML classifier that can automatically assign labels to LLM responses given an input claim, similar to prior efforts in stance detection and general question answering tasks~\cite{zhang2022would,aldayel2021stance,lin-etal-2022-truthfulqa}.
Third, introducing an automated classifier would increase the feasibility of extending our experiments to additional State-of-the-art LLMs such as Claude~\cite{claude} or LLaMa~\cite{llama}. 

Similarly, further experiments can be conducted to verify hypotheses surrounding LLM correctness in responding to varying S\&P categories. 
To illustrate, experiments can be designed to test whether complex legal language influences LLM performance among law and regulation misconceptions.  
Experiments can be repeated after substituting legal language in these misconceptions with plain language (commonly used among lay people).  

Future research should also experiment with prompt engineering. 
Instead of static prompts, more diverse prompt structures, such as chain-of-thought prompting~\cite{wei2022chain} and the ReAct prompt framework~\cite{yao2023react}, can be used to discover optimized prompts that can solicit correct output.
Prompts with varying levels of context can be used to discover the influence of providing more context (\eg \textit{``I heard a different opinion from a reputable source?''}) or nudging/challenging the LLM (\eg \textit{``are you sure about your response?''}) 

\shortsectionBf{Importance of Understanding LLM Tool Use.}
Users have increasingly shown a keen interest in leveraging LLM tools. 
Consider a scenario where a user seeks to confirm the correctness of a claim that is actually false. 
A single trial may incorrectly state that the claim is true (our results in \texttt{E2} show how LLMs can be inconsistent).
A user who does not repeatedly ask the question would never read the correct version of the response.
Our study shows the potential shortcomings of LLMs when used by end users. 
However, very little is known about \textit{how} users interact with newly-emerging LLM tools.
For instance, would users be inclined to query LLM tools repeatedly to ensure the veracity of the LLM's response? 
Despite a demonstrable lack of reliability in providing S\&P advice, assuming that users leverage tools with caution would be overoptimistic.
Thus, more user-centered research is required to understand user interactions with LLMs. 
Future work should explore qualitative efforts to understand (1)~how and (2)~when users interact with LLM interfaces. 
For instance, when seeking expert advice, do users interact with LLMs as a sole source of information or a preliminary step before verifying through other means (\eg asking friends/ family, referring to scientific papers)?

However, large-scale quantitative studies - (\eg surveys) are required for a deeper understanding of user-LLM interaction.
To illustrate, contextual factors such as the severity of the misconception may influence user interaction and trust of LLM  - \eg users are likely to have different trust levels when casually querying about the S\&P of IoT devices when compared to asking critical questions about the strengths of their password.
Understanding these factors is imperative for the development of LLM tools. 
To illustrate, the scenario in which users are inclined to trust LLMs as a sole source warrants prioritizing performance in providing valid URLs.

\subsection{Limitations}
\label{subsec:limitation}
Our study has three main limitations.  
First, S\&P misconceptions may differ between different demographics (\eg country of origin, socioeconomic status). 
We rely on existing manuscripts to inform us of publicly held user S\&P misconceptions, and thus, our dataset likely excludes misconceptions that are not publicly studied.
Given that such misconceptions are not publicly available, it is highly likely that LLM models trained on public information may not respond correctly to such misconceptions.
This can affect the overall correctness and model error rates.
Future work should focus on large-scale surveys targeting demographically diverse participants to generate representative datasets of S\&P misconceptions. 

Second, we focused on querying S\&P misconceptions in English. 
Performance can vary across languages, further influencing model effectiveness in refuting S\&P misconceptions and providing non-English sources of information.
Additional work is required to compare LLM reliability as an advice tool in different languages.

Finally, we use a static prompt template. % to query LLMs.
It is known that LLM responses could vary depending on the prompts. 
Fine-crafted prompts have the potential to enhance the quality of LLM responses.
However, given our prompts were designed to mimic end users who query state-of-the-art LLMs, we leave the impact of prompt engineering on S\&P advice to future work (as detailed in Section~\ref{subsec:lessons}).

%% file: 0-ACSAC-Minor-Revision/Content/related_work.tex
\section{Related work}
\shortsectionBf{LLM Performance in Question Answering.} 
LLMs' impressive ability to generate natural language and capability in tasks such as text completion and question answering warrant recent research efforts focused on evaluating their performance. % in these domains.
Most closely related to our study is the growing body of work that aims to understand the accuracy of LLMs in question-answering tasks.
A recent study introduced a benchmark called TruthfulQA to assess language models' truthfulness in generating answers to questions~\cite{lin-etal-2022-truthfulqa}. 
This dataset was designed to measure imitative falsehoods by including questions that people often answer incorrectly due to false beliefs or misconceptions. 
The authors tested OpenAI GPT family models and found that the largest models were generally the least truthful. 

Similarly, Bang et al.~\cite{bang2023multitask} have evaluated ChatGPT's ability to generate factual content and detect misinformation using the TruthfulQA dataset. 
They revealed that ChatGPT answered falsely to one-third of the questions in the TruthfulQA dataset designed to elicit imitative falsehoods.
Another recent work has extensively evaluated ChatGPT's reliability in generic question answering and its ability to identify unanswerable questions using $10$ open QA benchmarks across eight knowledge domains, including history, law, and recreation~\cite{shen2023chatgpt}.
They discovered the deficiency of ChatGPT in identifying unanswerable questions and its varying ability among different knowledge domains.
These works broadly explore question-answering instead of investigating LLMs use cases for specific/specialized domains, which can vary significantly depending on the training data of state-of-the-art LLM interfaces. 

In contrast, we curate a novel dataset of S\&P misconceptions after an extensive literature survey on user perception of S\&P technologies.
Interestingly, given that the misconceptions we collect are publicly held, they could potentially have influenced training data (since LLMs leverage large volumes of internet data). 
We then measure LLM's ability to refute popular misconceptions. 
\shortsectionBf{LLMs for Specialized Contexts.} Complementing research on LLM question-answering ability, research evaluating LLMs' capability in a specialized context (producing domain-specific content) has also been explored, primarily in health and medical-related fields.  
For instance, Zuccon et al.~\cite{zuccon2023dr} have evaluated ChatGPT's ability to answer complex health information questions and how knowledge provided in the prompt affects the accuracy of its answers. 
This work used various topics from a trusted resource (TREC Health Misinformation website) and showed that ChatGPT effectively answers health-related questions and debunks misconceptions about health treatment.
However, other research~\cite{birkun2023large,saeidnia2023evaluation,huang2023fact} note the shortcomings of LLMs in providing expert medical-related opinions, noting inaccurate/false generated content.  

In contrast, our work not only shifts focus to the S\&P domain, but we designed experiments to measure consistency, the influence of other factors (\eg paraphrasing), and LLM's ability to provide reliable sources. 
Our results not only confirm prior findings on the lack of reliability of state-of-the-art LLMs in providing specialized opinions but also extend findings by highlighting vulnerabilities to consistency, paraphrasing, and lack of reliable sources. 

\shortsectionBf{Limitations of LLM Performance.} 
Investigations into question-answering ability extend beyond measuring accuracy.
Recent works seek to understand LLM \textit{limitations} in answering questions or responding to user input. 
For instance,  prior research has evaluated their vulnerability in producing hallucinations~\cite{bang2023multitask, guerreiro2023hallucinations,zhang2023siren}, the tendency to generate falsehoods that appear true but are nonsensical (\eg referencing fake academic papers or news articles but claiming them to be true).
Related to this, Ren et al.~\cite{ren2023investigating} investigate whether LLMs are able to perceive their knowledge boundary (and acknowledge when they are not able to answer accurately). 
Our findings extend insight into LLM limitations by exposing LLM vulnerability to repeated queries and paraphrasing. 
Invalid URLs (from \texttt{E4}) underscore the prevalence of hallucination.
However, we also discover that LLMs can hallucinate by synthesizing URLs that point to an existing website but are unrelated to the question/prompt provided. 
This suggests that LLMs may ``hallucinate'' that an existing website contains domain-relevant information when, in reality, they do not. 
We also extend findings by showing that even with real-time web search capabilities, Bard may still produce invalid URLs.

\shortsectionBf{Using LLMs for Cybersecurity.} A growing body of work also investigates LLM from a cybersecurity perspective.
Broadly, these works explore LLMs' ability to generate bug-free code or to detect S\&P issues in vulnerable code~\cite{deng2023pentestgpt,pearce2023examining,caner2023zeroleak,paria2023divas,noever2023can}. 
These efforts help investigate the feasibility of LLMs to generate reliable code and also its use in an end-to-end software development pipeline.

In contrast, our research focuses on end-user interaction with cybersecurity (or, more broadly, Security \& Privacy), where we curate a novel S\&P dataset to understand LLM reliability in helping end users receive factually correct S\&P advice. 

\shortsectionBf{Security and Privacy Advice.} 
Existing studies on S\&P advice have targeted understanding of where and how users receive advice (\eg online forums, TV, peers, or through their own negative experiences) as well as its content and quality~\cite{nicholson2019if, redmiles2016think, rader2012stories}. 
Prior work has also studied how demographics (\eg socioeconomic status, skill level) impact resources for S\&P advice~\cite{redmiles2016learned, redmiles2017digital}. 
For instance, Redmiles et al.~\cite{redmiles2020comprehensive} conducted a comprehensive quality evaluation of the S\&P advice on the web and found that most users believed online advice was somewhat actionable and comprehensive. 
%S\&P
Another line of work has proposed dissemination methods, leveraging interventions such as interactive games and comic strips~\cite{denning2013control, srikwan2008using, zhang2016role}. 

In contrast, we focus on assessing emerging LLM user interfaces of ChatGPT and Bard. 
Given their increasing presence as a trusted information resource, we evaluate their ability to provide S\&P advice by refuting common S\&P misconceptions.

%% file: 0-ACSAC-Minor-Revision/Content/conclusion.tex
\section{Conclusions}
LLMs like ChatGPT and Bard have made prominent advancements in generative AI, becoming part of everyday users' lives.
Their growing function as a trusted information source warrants evaluating their ability to provide expert advice. 
We evaluate LLMs' ability to provide S\&P advice by refuting user-held S\&P misconceptions.
We first curate a dataset of $122$ S\&P misconceptions and query two popular LLMs in four experiments to measure overall correctness, consistency, and susceptibility to paraphrasing. 
We also analyze the sources LLMs provide to justify their stance towards misconceptions. 
We find that LLMs demonstrate non-negligible error rates, which increase when misconceptions are queried repeatedly or paraphrased. 
LLMs may show inconsistency, demonstrating multiple stances for a single misconception.
LLMs often provide invalid URLs as resources and, in cases of valid URLs, may erroneously refer to websites with irrelevant information.
Our work highlights LLM shortcomings in their reliability as an S\&P advice tool. 